\documentclass[iop]{emulateapj}
\usepackage{graphics}
\usepackage{amsmath}

\begin{document}

\title{Two timescale dispersal of magnetized protoplanetary disks}

\author{Philip J. Armitage\altaffilmark{1,2}, Jacob B. Simon\altaffilmark{1}, and Rebecca G. Martin\altaffilmark{1,3}}

\email{pja@jilau1.colorado.edu}

\begin{abstract}
Protoplanetary disks are likely to be threaded by a weak net flux of vertical magnetic field that is a remnant 
of the much larger fluxes present in molecular cloud cores. If this flux is approximately conserved its 
dynamical importance will increase as mass is accreted, 
initially by stimulating magnetorotational disk turbulence and subsequently by enabling wind angular momentum 
loss. We use fits to numerical simulations of ambipolar dominated disk turbulence to construct simplified 
one dimensional evolution models for weakly magnetized protoplanetary disks. We show that the late onset 
of significant angular momentum loss in a wind can give rise to ``two timescale" disk evolution in which a long 
phase of viscous evolution precedes rapid dispersal as the wind becomes dominant. The wide 
dispersion in disk lifetimes could therefore be due to varying initial levels of net flux. Magnetohydrodynamic (MHD)  
wind triggered dispersal differs from photoevaporative dispersal in predicting mass loss from 
small ($< 1 \ {\rm AU}$) scales, where thermal winds are suppressed. Our specific models are based on 
a limited set of simulations that remain uncertain, but qualitatively 
similar evolution appears likely if mass is lost from disks more quickly than flux, and  
if MHD winds become important as the plasma $\beta$ decreases.
\end{abstract} 


\altaffiltext{1}{JILA, University of Colorado and NIST, 440 UCB, Boulder, CO 80309-0440}
\altaffiltext{2}{Department of Astrophysical and Planetary Sciences, University of Colorado, Boulder}
\altaffiltext{3}{Sagan Fellow}

\section{Introduction} 
Protoplanetary disks have typical lifetimes of a few to 10 Myr \citep{haisch01,hernandez07,bell13}, but are dispersed 
on an order of magnitude shorter timescale \citep{simon95,wolk96}. This two timescale behavior is 
one of the basic observed properties of protoplanetary disks \citep{luhman10}, and is inconsistent with the predicted 
power-law decline in the surface density of a simple viscous accretion disk \citep{lyndenbell74}.  
It implies that there is a physically distinct dispersal process that rapidly removes the gas and 
dust at the end of the disk phase.

Photoevaporation leads to mass loss from disks exposed to 
ultraviolet and X-ray radiation \citep{alexander08,clarke11}. Since the work 
of \citet{clarke01}, who showed that photoevaporation leads to two timescale disk evolution, it has been 
the leading candidate dispersal mechanism. Here, we propose another. We assume that 
angular momentum transport  is a consequence of magnetohydrodynamic (MHD) 
turbulence, and that disks are threaded by a conserved weak vertical magnetic flux whose 
strength is a byproduct of the star formation process. 
As the disk accretes, the relative importance of the magnetic field 
(characterized by the ratio of magnetic to gas pressure at the disk mid-plane) increases, stimulating 
stronger MHD turbulence \citep{hawley95} and eventually angular momentum loss in a magnetized 
disk wind \citep{fromang13,bai13}. The presence of MHD disk winds has been shown to 
result in significant changes to the predicted structure of protoplanetary disks on AU scales 
\citep{bai13b}. We show that they may also lead to two timescale disk evolution resembling 
that produced by photoevaporation, but driven by the transport and loss of angular momentum 
rather than by mass loss. For 
the purposes of demonstrating the essential elements of our model we ignore the effects of 
{\em mass loss} from MHD disk winds (as well as from photoevaporation), along with 
potential couplings between MHD 
and thermal outflows \citep[akin to the radiation / MHD winds discussed by][]{proga03}. Mass loss, 
of course, would occur in a complete MHD wind model, and could 
in itself drive dispersal \citep{suzuki09}. Depending upon the radial distribution of mass loss, 
its effects could be almost indistinguishable from photoevaporation.

We detail our one-dimensional disk model in \S2. The key inputs are the transport and loss of 
angular momentum in the outer disk, where the bulk of the mass resides and 
where ambipolar diffusion is important \citep{armitage11}. We use results from 
simulations by \citet{simon13a,simon13b} to evaluate these quantities, 
while acknowledging that there are large uncertainties due to the local nature of these calculations.
We show results for the long term disk evolution in \S3, and discuss 
the implications of MHD-driven disk dispersal in \S4.

\section{Disk model}
We model disk evolution using a one dimensional vertically integrated model 
that includes internal redistribution of angular momentum (``viscosity") and wind angular momentum loss. 
The model is similar in spirit to one for black hole disk variability proposed by 
\citet{king04}. 
We assume a fixed mid-plane temperature profile $T \propto r^{-1/2}$ appropriate for 
disk evolution at late times and large radii, when viscous heating is negligible and the thermal 
balance is dominated by stellar irradiation \citep{kenyon87}. For a mid-plane sound speed $c_s$ the 
disk scale height $h = c_s / \Omega$, where $\Omega$ is the Keplerian angular velocity. We take,
\begin{equation}
 \frac{h}{r} = 0.05 \left( \frac{r}{10 \ {\rm AU}} \right)^{1/4} .
\end{equation}
Adopting a Shakura-Sunyaev (1973) form for the viscosity, $\nu = \alpha h^2 \Omega$, with a 
constant $\alpha$ (which we do {\em not} assume later), the thermal structure of the disk 
implies $\nu \propto r$. For the initial surface density we take a steady state radial 
profile truncated with an exponential cut-off. In the constant $\alpha$ case this is just 
the usual similarity solution \citep{lyndenbell74},
\begin{equation}
 \Sigma (t=0) \propto r^{-1} \exp (-r / r_0).
\label{eq_initial} 
\end{equation}
For the calculations presented in $\S 3$ we take 
$r_0 = 10 \ {\rm AU}$ and set the surface density normalization such that the initial disk mass is $0.03 \ M_\odot$. 
Up to this point, the assumptions closely match those made in evolutionary models for 
dispersal via photoevaporation, which are consistent with observations when evolved with 
$\alpha \sim 10^{-3} - 10^{-2}$ and wind mass loss rates $\dot{M}_{\rm w} \sim 10^{-10} \ M_\odot \ 
{\rm yr}^{-1}$ or higher.

We evolve the surface density of the disk under the action of internal angular momentum redistribution 
via MHD turbulence \citep[approximately modeled as a local, viscous process,][]{balbus99,simon12} and external 
angular momentum loss in a wind. 
The ambipolar disk simulations show that the internal stress develops a substantial laminar component when the 
net field is strong, but we treat all internal stresses as if they were a local viscosity.  
Neglecting any sources of mass loss other than accretion we 
have,
\begin{equation} 
 \frac{\partial \Sigma}{\partial t} = 
 \frac{3}{r} \frac{\partial}{\partial r} \left[ r^{1/2} \frac{\partial}{\partial r} \left( \nu \Sigma r^{1/2} \right) \right]
 - \frac{1}{r} \frac{\partial}{\partial r} \left( r \Sigma v_r \right),
\label{eq_evolve} 
\end{equation}
where $v_r$ is the radial velocity induced by the wind. Numerical simulations show that both $\nu$ 
(or, more usefully, $\alpha$) and $v_r$ are functions of the vertical magnetic field $B_z$ that locally threads the 
disk. We parameterize this field via the ratio of the mid-plane thermal and magnetic pressures,
\begin{equation}
 \beta_z = 8 \pi \frac{\rho c_s^2}{B_z^2},
\end{equation}
where the mid-plane density $\rho = (1 / \sqrt{2 \pi}) (\Sigma / h)$. How the disk evolves depends on 
the functional forms of $\alpha(\beta_z)$ and $v_r (\beta_z)$, and on how $B_z(r,t)$ varies 
as gas is accreted.

\begin{figure}
\begin{center}
\includegraphics[width=0.45\textwidth,angle=0]{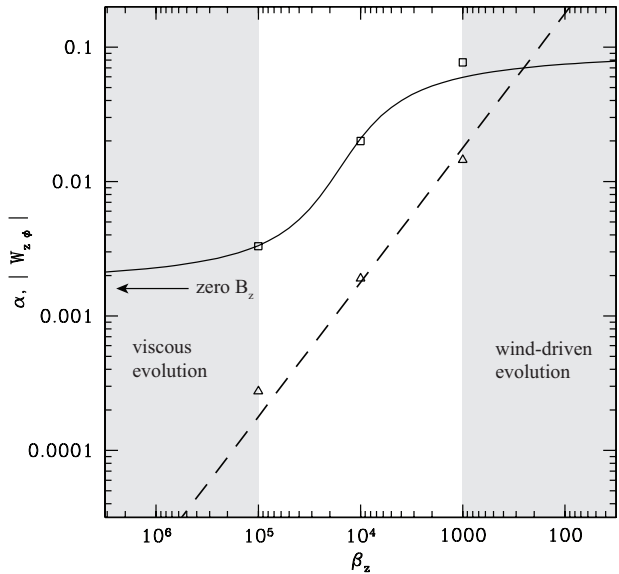}
\end{center}
\vspace{-0.2truein}
\caption{Measurements of how the internal disk stress $\alpha$ (squares) and the wind stress 
$| \overline{W_{z \phi}} |$ (triangles) vary with the vertical magnetic field $\beta_z$, based on 
numerical simulations of the ambipolar MRI at 30~AU \citep{simon13b}. The  lines show 
fitting formulae used in the simplified disk models. The stresses are 
not directly comparable (the wind stress measured in this manner is more efficient at driving 
inflow by a factor $\sim (h/r)^{-1}$), but the results imply that wind angular momentum loss 
becomes significant for disk evolution for $\beta_z \lesssim 10^4$. The arrow shows an upper limit 
to the stress without any net field \citep{simon13a}.}
\vspace{0.1truein}
\label{fig1}
\end{figure}

A vertical magnetic field influences the dynamics of 
the disk in two ways. First, stronger fields increase the strength of turbulent transport \citep{hawley95}. 
Second, sufficiently strong fields lead to the formation of an MHD disk wind which can carry away 
both mass and angular momentum \citep{blandford82}. These effects appear to be generic for 
disks that are unstable to the magnetorotational instability \citep[MRI,][]{balbus98}, though the details 
of how the turbulence and wind vary with the field strength depend on the included disk 
physics \citep{suzuki09,fromang13,bai13}. We are interested in the long term  
evolution of the disk, so what matters most is the MRI disk physics on the large scales where 
most of the mass resides and where the viscous timescale is the longest. Here, ambipolar 
diffusion is an important non-ideal MHD effect \citep{kunz04,bai11,perezbecker11}. 
Information as to the strength of turbulence and winds in this regime is available from the 
vertically stratified ambipolar disk simulations of \citet{simon13b}, who model MRI 
turbulence at $30 \ {\rm AU}$ in a disk ionized by stellar FUV photons. We characterize the 
wind stress via a dimensionless parameter $| \overline{W_{z \phi}} |$ 
calculated directly from the numerical simulations (we use the stress derived from an estimate of where 
the base of the wind lies). $| \overline{W_{z \phi}} |$ is related to the 
induced radial velocity via,
\begin{equation}
 v_r = - \frac{4}{\sqrt{2 \pi}} \vert \overline{W_{z \phi}} \vert c_s.
\end{equation} 
Figure~\ref{fig1} shows the measurements of $\alpha(\beta_z)$ and $| \overline{W_{z \phi}} | (\beta_z)$, 
along with fitting formulae used in our one dimensional disk models. We adopt,
\begin{eqnarray}
 \log \alpha & = & A + B {\rm tan}^{-1} \left[ \frac{C - \log \beta_z}{D} \right] \nonumber \\
 \log \vert \overline{W_{z \phi}} \vert & = & 1.25 - \log \beta_z ,
\end{eqnarray} 
with $A=-1.9$, $B=0.57$, $C=4.2$ and $D=0.5$. Obviously, the numerical data 
to which this fit is anchored is sparse, and the values themselves are uncertain. 
We have implicitly assumed a specific ionization model, and neglected the radial dependence of the  
stress across the ambipolar-dominated zone. The inner (Ohmic-dominated) disk physics would be 
significantly different, but should not affect the accretion rate evolution provided that a 
steady-state is established close to the star. The generic conclusions we draw from the simulations are two-fold: 
(i) internal disk transport 
is stimulated by a vertical field, and (ii) for strong enough fields angular momentum loss in a wind controls disk evolution. 
The first of these is robust but of limited 
importance (our model would work equally well with a constant $\alpha$), the second 
is the critical ingredient.

The disk evolution also depends on how the magnetic flux evolves 
as gas accretes. Many complexities lurk here. Magnetic flux can be dragged 
inward by the mean flow, diffuse as a byproduct of turbulence \citep{lubow94a,guilet13}, and be lost 
from the disk at its radial boundaries. Unfortunately 
the ratio of the turbulent diffusion to the turbulent viscosity remains uncertain in ideal MHD 
and is unknown in the ambipolar dominated limit. The simplest model assumes 
that the magnetic flux,
\begin{equation}
 \Phi = \int_{r=r_{\rm in}}^{r(\Sigma=\Sigma_{\rm small})} 2 \pi r B_z {\rm d}r ,
\end{equation}
is conserved {\em within the disk}, whose outer edge we define at 
$\Sigma_{\rm small} = 10^{-2} \ {\rm g \ cm}^{-2}$. 
We further assume for most of our models 
that diffusion acts to keep $\beta_z$ 
spatially constant as the disk evolves. Together with the fixed thermal structure, these 
assumptions specify the time evolution of $\beta_z$ given an initial magnetic flux and 
the evolution of $\Sigma(r)$.

There is an observational argument that flux is not all lost from the disk 
inner edge. Ordered T~Tauri magnetic fields \citep{yang11} are strong in absolute 
terms ($B \approx {\rm kG}$) but weaker than they would be if any significant 
disk flux was accreted (e.g. $\beta_z = 10^5$ at 10~AU in our initial models 
implies $B_z \simeq 5 \ {\rm mG}$, which would lead to a stellar field of 
$\approx 20 \ {\rm kG}$ if accreted). Flux probably is, however, lost from the 
outer disk. 

We solve Eq.~\ref{eq_evolve} with an explicit finite difference scheme on a 
logarithmic grid, with 300 zones between $r_{\rm in} = 0.067 \ {\rm AU}$ and 
$6. 7 \times 10^3 \ {\rm AU}$. Zero torque boundary conditions are applied at  
both boundaries.

\section{Results}

\begin{figure}
\begin{center}
\includegraphics[width=0.45\textwidth,angle=0]{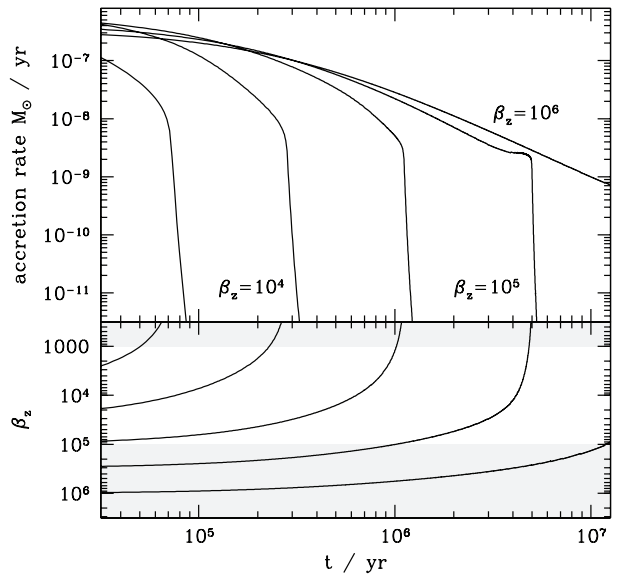}
\end{center}
\vspace{-0.2truein}
\caption{The predicted evolution of the accretion rate into the inner disk $\dot{M}$ (upper panel) and 
vertical field strength parameter $\beta_z$ (lower panel, assumed spatially uniform 
across the disk), for initial $\beta_z$ equal to $10^4$, $3 \times 10^4$, $10^5$, 
$3 \times 10^5$, and $10^6$. Two timescale disk evolution, with dispersal within 
10~Myr, is predicted to occur for $3 \times 10^4 \le \beta_z \le 3 \times 10^5$. Stronger fields 
lead to short lifetimes, fields weaker than this range result in an extended period of purely 
viscous power-law evolution.}
\label{fig2}
\end{figure}

Figure~\ref{fig2} shows the time evolution of the 
inner accretion rate $\dot{M}$ for varying initial $\beta_z$ and exponentially truncated 
steady-state disk surface density profiles. The uppermost curve shows the predicted 
evolution for an initial $\beta_z = 10^6$, which is high enough that the disk remains 
in almost the zero net field limit for more than $10^7$~yr. The resulting 
$\alpha$ is almost constant, there is no wind, and the evolution closely approximates 
the similarity solution. No two timescale behavior is seen. Stronger initial fields 
yield a clear transition from an initial viscous phase of slow evolution 
into a wind-driven phase and rapid dispersal. For a broad range of 
initial fields, $10^4 \le \beta_z \le 3 \times 10^5$, the transition to dispersal 
occurs for stellar accretion rates that lie between $10^{-9} \ M_\odot \ {\rm yr}^{-1}$ 
and $10^{-8} \ M_\odot \ {\rm yr}^{-1}$. This transition $\dot{M}$ lies in between 
those predicted by different internal photoevaporation models, which range 
between $\sim 10^{-10} \ M_\odot \ {\rm yr}^{-1}$ \citep[for diffuse EUV irradiation,][]{font04} 
and in excess of $\sim 10^{-8} \ M_\odot \ {\rm yr}^{-1}$ \citep[for X-ray or FUV 
driven flows,][]{gorti09,owen10}. The disk lifetimes 
range from less than 0.2~Myr for an initial $\beta_z = 10^4$ to more than 10~Myr 
for weak-field disks. How the predicted evolution of {\em individual} disks translates 
into the predicted evolution of the disk {\em population} depends on the assumed 
distribution of $\beta_z$, but if the typical value is such as to produce a few Myr 
disk lifetime then the population statistics would also provide evidence for two 
timescale evolution.

The two timescale evolution is a consequence of the way 
that the viscous and wind stresses extracted from the simulations vary with vertical 
field strength. For a starting $\beta_z \sim 10^5$, the initial evolution of the disk 
is dominated by viscous spreading and accretion, with the wind playing a minor 
role. As mass is unloaded from the vertical field due to accretion, flux 
conservation implies that $\beta_z$ decreases. (This is true despite the increasing 
disk area for the models considered here.) 
Initially the main effect of the relatively stronger field is to increase the internal 
stresses, yielding an $\dot{M} (t)$ curve that rolls over faster than it would if 
$\alpha$ were fixed. As $\beta_z$ continues to decrease, the loss of angular 
momentum in the wind starts to contribute and eventually dominates the disk 
evolution. The steep dependence of $| \overline{W_{z \phi}} |$ on $\beta_z$, 
together with the monotonic inflow occasioned by the advective nature of the 
wind term, leads to a runaway effect. The disk is dispersed rapidly entirely by accretion 
onto the star. We do not model the feedback effect that the disk inflow exerts on the 
wind structure, which may lead to instability when the wind becomes dominant \citep{lubow94b}. 
If this effect is present for protoplanetary disks, it would presumably speed dispersal further.

The measurements of wind stresses from local simulations (Figure~\ref{fig1}) are 
uncertain, because these simulations do not represent the global 
geometry that is a key aspect of winds \citep[e.g.][]{blandford82}. One might 
therefore ask whether the increase in $\alpha$ as the field becomes stronger might 
suffice to yield two timescale disk evolution, without involving 
the viscous-to-wind transition at all. It does not. Although the increase in internal stresses as the disk evolves 
substantially reduces the viscous timescale at fixed radius, the ongoing 
expansion of the disk means that there is no distinct dispersal phase and the 
late time evolution of $\dot{M}$ follows a power law.

\begin{figure}
\begin{center}
\includegraphics[width=0.45\textwidth,angle=0]{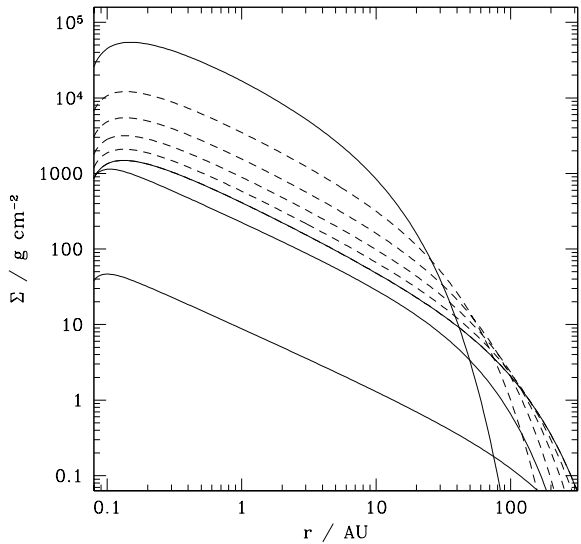}
\end{center}
\vspace{-0.2truein}
\caption{Evolution of the surface density for models with initial $\beta_z = 10^6$ (dashed curves) 
and $10^5$ (solid curves), starting from identical initial conditions (upper solid curve). The time 
slices are at equal intervals of 0.43~Myr. The onset of the wind reverses the viscous expansion 
of the disk, but the dispersal occurs approximately uniformly across all radii.}
\label{fig3}
\end{figure}

\begin{figure}
\begin{center}
\includegraphics[width=0.45\textwidth,angle=0]{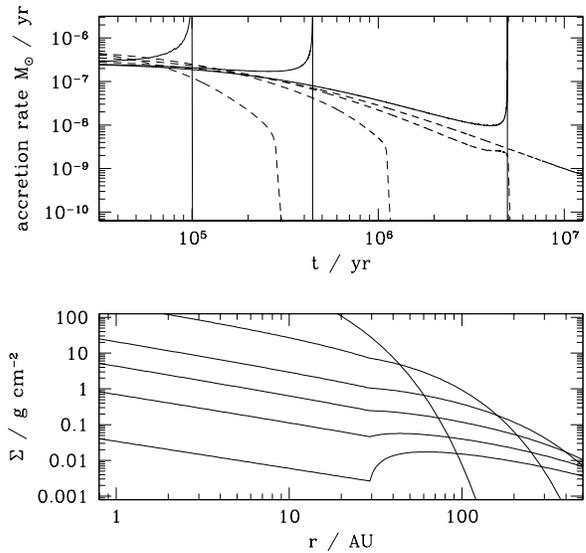}
\end{center}
\vspace{-0.2truein}
\caption{Disk evolution in models that vary the assumptions for the flux evolution and wind properties.
The upper panel shows the evolution of $\dot{M}$ for the fiducial models (dashed lines, corresponding 
to those in Figure~\ref{fig2}), along with models where turbulence is assumed to maintain constant 
differential flux per unit radius (solid lines). From left to right, the solid lines correspond to 
disks that have the same total flux as fiducial models with $\beta_z = 10^6, 3 \times 10^6, 
10^7$. The lower panel shows $\Sigma(r,t)$ for a fiducial model with $\beta_z = 10^5$, but where the wind 
is assumed not to operate outside of 30~AU.}
\label{fig4}
\end{figure}

Figure~\ref{fig3} shows the predicted evolution of the surface density with time for weakly and moderately 
magnetized disks. The onset of wind angular momentum loss in the moderately magnetized 
case prevents some of the viscous expansion that would otherwise occur\footnote{The evolution of 
disk radius with time is a diagnostic of the importance of internal stresses in dwarf novae disks \citep{smak84}.}, 
but in broad terms the surface density during wind-driven dispersal 
declines smoothly across all radii \citep[observationally, 
``homologously depleting disks",][]{wood02}. This behavior of our models, however, results  
from assuming that $\beta_z$ has no spatial dependence. It is not a general 
consequence of wind-driven dispersal. To illustrate this, we show in Figure~\ref{fig4} how 
the evolution differs if we vary elements of the model. First, we assume that instead of 
$\beta_z$ remaining constant with radius as the disk evolves, it is instead the differential 
flux ${\rm d} \Phi / {\rm d}r$ that remains fixed. This change weights the (fixed) 
flux toward the outer regions of the disk, which are therefore wind-dominated at all times. 
Dispersal in this case is extremely swift but outside-in. The reduction of the disk area with time means that 
the gas is finally swept into the star in an  
accretion burst. Second, we consider a model with the fiducial flux evolution, 
but switch off the action of the wind outside of an arbitrary radius which we take to be 
30~AU. In this case there is no two timescale evolution, as the purely viscous outer 
disk is not able to be dispersed by MHD processes. Rather, a cavity develops in the 
inner disk once the wind becomes important on small scales, with the cavity being fed by the viscous reservoir 
further out. For the specific case shown in the Figure (with initial $\beta_z = 10^5$), 
the hole only forms when the disk has a low mass, but this is again model-dependent.
Clearly, MHD-driven dispersal admits a range of phenomenology, only some of which is 
consistent with observations.  
 
\section{Discussion}
Large-scale magnetic fields could play a role in disk dispersal, 
even if they are too weak to prevent disk formation or to affect the early dynamics. 
This conclusion rests on several assumptions, (i) that disks are threaded by weak vertical 
fields, (ii) that there is a transition to a wind-dominated angular momentum loss regime 
as magnetic field pressure increases, and (iii) critically, that flux is lost from the outer disk more 
slowly than mass is accreted. The first 
assumption seems plausible given that one does not expect the strong fields present during 
star formation to be driven exactly to zero.  
The second is supported by numerical 
simulations \citep{suzuki09,fromang13,bai13,simon13b}, though these results  
are preliminary rather than definitive. Using  
a one dimensional disk model motivated by simulation results we found that two 
timescale disk evolution can occur, though it would 
be premature to assert that it is inevitable given our current knowledge of magnetic 
field evolution within disks. In a magnetic dispersal model the disk lifetime is 
tied to the strength of the initial magnetic field. Different 
field strengths can lead to prompt dispersal 
($\ll 1 \ {\rm Myr}$) or to disks that are almost inviscid and surprisingly 
massive at late times \citep{bergin13}. 

We cannot say whether MHD dispersal is consistent with other observed 
properties of disk dispersal, such as the morphology of transition disks or the evidence 
that dispersal commonly occurs from the inside out \citep{koepferl13}. Whether such 
behavior occurs in MHD dispersal models depends upon how the radial evolution 
of the net field couples to the disk evolution. If $\beta_z$ first runs away into the dispersal 
phase in the inner disk the outcome is a cavity, whereas an outside-in runaway occurs if 
the flux distribution favors a strong MHD wind from the outer disk. 
Additional simulations will be required to decide 
whether non-homologous dispersal is a consequence of non-ideal MRI disk physics, 
and to determine how rapidly flux can escape from evolving disks entirely.

Although the mass loss rates from local MHD winds are poorly determined, 
the radial extent of the wind is expected to be broad and to include the inner 
disk \citep{bai13c}. The key contrast to photoevaporative outflows --- which fall off rapidly 
in strength inside a critical radius $r_c \approx 0.2 GM_* / c_s^2 \approx 2 \ {\rm AU}$ 
(where $c_s$ is the sound speed in the heated gas) --- is thus that MHD winds would be expected to 
include a higher velocity component launched from closer to the star.

\acknowledgments

We thank our collaborator Xuening Bai, Richard Alexander and Cathie Clarke for help and advice.  
PJA and  JBS acknowledge support from NASA grants NNX11AE12G and NNX13AI58G, and from 
grant HST-AR-12814 awarded by the Space Telescope Science Institute, which is operated by the 
Association of Universities for Research in Astronomy, Inc., for NASA, under contact NAS 5-26555. 
RGM's support was provided in part under contract with the
California Institute of Technology (Caltech) funded by NASA
through the Sagan Fellowship Program.


\begin{thebibliography}{62}
\expandafter\ifx\csname natexlab\endcsname\relax\def\natexlab#1{#1}\fi 

\bibitem[Alexander(2008)]{alexander08}
Alexander, R. 2008, New Astronomy Reviews, 52, 60

\bibitem[Armitage(2011)]{armitage11}
Armitage, P.~J. 2011, ARA\&A, 49, 195

\bibitem[Balbus \& Papaloizou(1999)]{balbus99}
Balbus, S. A., \& Papaloizou, J. C. B. 1999, ApJ, 521, 650

\bibitem[Bai(2011)]{bai11}
Bai, X.-N. 2011, ApJ, 739, 50

\bibitem[Bai(2013)]{bai13b}
Bai, X.-N. 2013, ApJ, 772, article id. 96

\bibitem[Bai \& Stone(2013)]{bai13}
Bai, X.-N., \& Stone, J. M. 2013, ApJ, 767, 30

\bibitem[Bai \& Stone(2013b)]{bai13c}
Bai, X.-N., \& Stone, J. M. 2013b, ApJ, 769, 76

\bibitem[Balbus \& Hawley(1998)]{balbus98}
Balbus, S. A., \& Hawley, J. F. 1998, Rev. Mod. Phys., 70, 1

\bibitem[Bell et al.(2013)]{bell13}
Bell, C. P. M., Naylor, T., Mayne, N. J., Jeffries, R. D., \& Littlefair, S. P. 2013, MNRAS, 434, 806

\bibitem[Bergin et al.(2013)]{bergin13}
Bergin, E. A., et al. 2013, Nature, 493, 644

\bibitem[Blandford \& Payne(1982)]{blandford82}
Blandford, R. D., \& Payne, D. G. 1982, MNRAS, 199, 883

\bibitem[Clarke(2011)]{clarke11}
Clarke, C. J. 2011, in Physical Processes in Circumstellar Disks around Young Stars, ed. Paulo J.V. Garcia. IUniversity of Chicago Press, p.~355

\bibitem[{Clarke et al.(2001)}]{clarke01}
Clarke, C. J., Gendrin, A., \& Sotomayor, M. 2001, MNRAS, 328, 485

\bibitem[Font et al.(2004)]{font04}
Font, A. S., McCarthy, I. G., Johnstone, D., \& Ballantyne, D. R. 2004, ApJ, 607, 890

\bibitem[Fromang et al.(2013)]{fromang13}
Fromang, S., Latter, H., Lesur, G., \& Ogilvie, G. I. 2013, A\&A, 552, id.~A71

\bibitem[Gorti \& Hollenbach(2009)]{gorti09}
Gorti, U., \& Hollenbach, D. 2009, ApJ, 690, 1539

\bibitem[Guilet \& Ogilvie(2013)]{guilet13}
Guilet, J., \& Ogilvie, G. I. 2013, MNRAS, 430, 822

\bibitem[{Haisch et al.(2001)}]{haisch01}
Haisch, K. E., Jr., Lada, E. A., \& Lada, C. J. 2001, ApJ, 553, L153

\bibitem[Hawley et al.(1995)]{hawley95}
Hawley, J. F., Gammie, C. F., \& Balbus, S. A. 1995, ApJ, 440, 742

\bibitem[{Hern\'andez et al.(2007)}]{hernandez07}
Hern\'andez, J., et al. 2007, ApJ, 662, 1067

\bibitem[Kenyon \& Hartmann(1987)]{kenyon87}
Kenyon, S. J., \& Hartmann, L. 1987, ApJ, 323, 714

\bibitem[King et al.(2004)]{king04}
King, A. R., Pringle, J. E., West, R. G., \& Livio, M. 2004, MNRAS, 348, 111

\bibitem[Koepferl et al.(2013)]{koepferl13}
Koepferl, C. M., Ercolano, B., Dale, J., Teixeira, P. S., Ratzka, T., \& Spezzi, L. 2013, MNRAS, 428, 3327

\bibitem[Kunz \& Balbus(2004)]{kunz04}
Kunz, M. W., Balbus, S. A. 2004, MNRAS, 348, 355

\bibitem[Lubow et al.(1994a)]{lubow94a}
Lubow, S. H., Papaloizou, J. C. B., \& Pringle, J. E. 1994, MNRAS, 267, 235

\bibitem[Lubow et al.(1994b)]{lubow94b}
Lubow, S. H., Papaloizou, J. C. B., \& Pringle, J. E. 1994, MNRAS, 268, 1010

\bibitem[Luhman et al.(2010)]{luhman10}
Luhman, K. L., Allen, P. R., Espaillat, C., Hartmann, L., \& Calvet, N. 2010,ÊApJS, 186, 111

\bibitem[Lynden-Bell \& Pringle(1974)]{lyndenbell74}
Lynden-Bell, D., \& Pringle, J. E. 1974, MNRAS, 168, 603

\bibitem[Owen et al.(2010)]{owen10}
Owen, J. E., Ercolano, B., Clarke, C. J., \& Alexander, R. D. 2010, MNRAS, 401, 1415

\bibitem[Perez-Becker \& Chiang(2011)]{perezbecker11}
Perez-Becker, D., \& Chiang, E. 2011, ApJ, 735, 8

\bibitem[Proga(2003)]{proga03}
Proga, D. 2003, ApJ, 585, 406

\bibitem[Shakura \& Sunyaev(1973)]{shakura73}
Shakura, N. I., \& Sunyaev, R. A. 1973, A\&A, 24, 337

\bibitem[Simon et al.(2013b)]{simon13b}
Simon, J. B., Bai, X.-N., Armitage, P. J., Stone, J. M., \& Beckwith, K. 2013, ApJ, 775, article id. 73

\bibitem[Simon et al.(2013a)]{simon13a}
Simon, J. B., Bai, X.-N., Stone, J. M., Armitage, P. J., \& Beckwith, K. 2013, ApJ, 764, article id. 66

\bibitem[Simon et al.(2012)]{simon12}
Simon, J. B., Beckwith, K., \& Armitage, P. J. 2012, MNRAS, 422, 2685

\bibitem[Simon \& Prato(1995)]{simon95}
Simon, M., \& Prato, L. 1995, ApJ, 450, 824

\bibitem[Smak(1984)]{smak84}
Smak, J. 1984, Acta Astronomica, 34, 161

\bibitem[Suzuki \& Inutsuka(2009)]{suzuki09}
Suzuki, T. K., Inutsuka, S. 2009, ApJ, 691, L49

\bibitem[Wolk \& Walter(1996)]{wolk96}
Wolk, S. J., \& Walter, F. M. 1996, AJ, 111, 2066

\bibitem[Wood et al.(2002)]{wood02}
Wood, K., Lada, C. J., Bjorkman, J. E., Kenyon, S. J., Whitney, B., \& Wolff, M. J. 2002, ApJ, 567, 1183

\bibitem[Yang \& Johns-Krull(2011)]{yang11}
Yang, H., \& Johns-Krull, C. M. 2011, ApJ, 729, 83

\end{thebibliography}
\end{document}